\documentclass[a4paper]{article}

\usepackage[utf8]{inputenc}
\usepackage{authblk}
\usepackage{amsmath}
\usepackage{amsthm}
\usepackage{bbold}
\usepackage{enumitem}
\usepackage[autosize]{dot2texi}
\usepackage{tikz}
\usetikzlibrary{shapes,arrows}
\usepackage{subcaption}
\usepackage{xcolor}
\usepackage[toc]{appendix}
\usepackage{hyperref}

\title{ABCDE: Application-Based Cluster Diff Evals}
\author[1]{Stephan van Staden}
\author[1]{Alexander Grubb}
\affil[1]{Google LLC}
\date{July 2024}

\newcommand{\Base}{\mathit{Base}}
\newcommand{\Exp}{\mathit{Exp}}

\newcommand{\weight}{\mathit{weight}}
\newcommand{\JaccardDistance}{\mathit{JaccardDistance}}

\newcommand{\SplitRate}{\mathit{SplitRate}}
\newcommand{\MergeRate}{\mathit{MergeRate}}
\newcommand{\Precision}{\mathit{Precision}}
\newcommand{\Recall}{\mathit{Recall}}
\newcommand{\OverMergeRate}{\mathit{OverMergeRate}}
\newcommand{\UnderMergeRate}{\mathit{UnderMergeRate}}
\newcommand{\GoodSplitRate}{\mathit{GoodSplitRate}}
\newcommand{\BadSplitRate}{\mathit{BadSplitRate}}
\newcommand{\GoodMergeRate}{\mathit{GoodMergeRate}}
\newcommand{\BadMergeRate}{\mathit{BadMergeRate}}
\newcommand{\GoodBadSplitMergeRate}{\mathit{(Good|Bad)(Split|Merge)Rate}}
\newcommand{\DeltaPrecision}{\Delta\mathit{Precision}}
\newcommand{\DeltaPrecisionContribution}{\Delta\mathit{PrecisionContribution}}
\newcommand{\indicate}{\mathbb{1}}
\newcommand{\pairweight}{\mathit{pairweight}}
\newcommand{\match}{\equiv}
\newcommand{\distinct}{\not\equiv}

\begin{document}

\maketitle

\begin{abstract}
This paper considers the problem of evaluating clusterings of very large populations of items. Given two clusterings, namely a Baseline clustering and an Experiment clustering, the tasks are twofold: 1) characterize their differences, and 2) determine which clustering is better. ABCDE is a novel evaluation technique for accomplishing that. It aims to be practical: it allows items to have associated importance values that are application-specific, it is frugal in its use of human judgements when determining which clustering is better, and it can report metrics for arbitrary slices of items, thereby facilitating understanding and debugging. The approach to measuring the delta in the clustering quality is novel: instead of trying to construct an expensive ground truth up front and evaluating the each clustering with respect to that, where the ground truth must effectively pre-anticipate clustering changes, ABCDE samples questions for judgement on the basis of the actual diffs between the clusterings. ABCDE builds upon the pointwise metrics for clustering evaluation, which make the ABCDE metrics intuitive and simple to understand. The mathematical elegance of the pointwise metrics equip ABCDE with rigorous yet practical ways to explore the clustering diffs and to estimate the quality delta.
\end{abstract}


\section{Introduction}

Clustering is the partitioning of a set of items into separate groups, called clusters. The items in each cluster should typically be similar, while the items from different clusters should be different. Although that sounds simple, there are significant challenges in many practical applications of clustering. For example, the criteria of what makes items similar or different might be complex and require human judgement. A clustering algorithm must then approximate that somehow. Moreover, if billions of items must be clustered, then there are typically also constraints such as computing time and cost.

In such complex settings, there is often no optimal clustering algorithm, and developers can experiment with many ideas to improve the quality of the clustering while satisfying the resource constraints. Understanding the resource constraints of a run of an algorithm is fairly straightforward. Understanding the resulting clustering itself, in particular its quality, can be challenging. Yet that is very important for effective development and for the consumers of the clustering.

During development, a new clustering can be evaluated with respect to a ground truth clustering to get a sense of its quality. A ground truth clustering consists of a small set of items that humans partitioned into ideal clusters. Once the human work is finished, the ground truth clustering can be stored and used repeatedly to evaluate new clusterings. Such evaluations are fast and they provide developers with quick feedback. However, ground truth clusterings typically don't cover all corner cases and all slices of the item data, so their evaluation metrics might not extrapolate well to the whole population of items. For example, a ground truth clustering cannot foresee the kinds of overmerges (i.e. $\Precision$ losses) that future experiments will perform, and the overmerged items might not have existed yet when the ground truth clustering was created. Other more expensive evaluation techniques, such as ABCDE, can be used to fill that gap, but typically only after the developer is satisfied with the evaluation results of the ground truth clusterings.

ABCDE stands for ``Application-Based Cluster Diff Evals". It is application-based, because one can associate each item with an importance value, or weight, that can be tailored to the application at hand. It is an evaluation technique that characterizes the differences between two given clusterings, namely a Baseline clustering (henceforth referred to as $\Base$) and an Experiment clustering (henceforth referred to as $\Exp$). Each of these clusterings is a partitioning of the entire population of items. ABCDE characterizes the differences between $\Base$ and $\Exp$ with two sets of metrics:
\begin{itemize}
    \item \textit{Impact metrics} describe the relationship between $\Base$ and $\Exp$ without reference to quality. These metrics do not require human judgements; they can be calculated automatically and exactly.
    \item \textit{Quality metrics} describe the difference in quality between $\Base$ and $\Exp$. Their calculation involves the sampling of questions for humans and statistical estimation.
\end{itemize}

As mentioned before, a ground truth clustering describes the ideal situation for a small set of items that have been selected in advance. It serves as a (very partial) specification: its quality metrics characterize the degree to which a large clustering, such as $\Exp$, satisfies the expectations it encodes. The quality metrics of ABCDE operate differently: ABCDE first considers the differences between $\Base$ and $\Exp$, which involve the whole population of items, before it samples questions for humans in order to estimate the difference in quality between the clusterings. One can think about it as a lazy on-demand expansion of the full ground truth that targets the items that experienced change and the nature of the changes. Doing that puts human judgements on the critical path, which means higher cost\footnote{Depending on the setting, it might be possible to reduce the amortized cost of ABCDE's human judgements: since they contain examples where either $\Base$ or $\Exp$ made a mistake, they can be collected and used to improve the clustering algorithm itself, or serve as a set of expectations for future clusterings.} and a longer waiting time before metrics are available, but the resulting assessment has high fidelity because it takes the entire clustering into account and it focuses on the actual clustering diffs (instead of pre-anticipated diffs).

ABCDE makes heavy use of the pointwise clustering metrics~\cite{vanstaden2024pointwise} in its impact metrics and its quality metrics. The pointwise metrics are intuitive and simple to understand. They characterize impact and quality aspects that are important for practical applications. It is straightforward to report impact metrics for arbitrary slices of items, which facilitates understanding and debugging. Moreover, the mathematical elegance of the pointwise metrics helps a lot to derive rigorous techniques for estimating the difference in quality.

\section{Clusterings and clustering algorithms}

Given a finite set of items $T$ and an equivalence relation $\match$ (i.e. a binary relation that is reflexive, symmetric and transitive), a \textit{cluster} is an equivalence class of $T$ with respect to $\match$, and a \textit{clustering} is the set of all clusters, i.e. a partitioning of $T$ into its equivalence classes.

In practical applications, the set of items $T$ can be very large and the ideal equivalence relation is not fully known. Humans can consider a pair of items and say whether they are truly equivalent or not, but since that does not scale to billions of items, we have only very sparse information about the ideal relation. The main job of a clustering algorithm in such a setting is to approximate the ideal equivalence relation. This is typically done by 1) deciding which items might be related (also called `blocking'), and 2) deciding which of these are equivalent according to a computable function that imitates the human judgements. The design space of clustering algorithms is consequently huge, and it becomes desirable to be able to evaluate the clustering results to determine which algorithm and configuration to prefer in practice.

So in the context of ABCDE, we have an approximated equivalence relation for $\Base$, and an approximated equivalence relation for $\Exp$. In fact $\Base$ and $\Exp$ might have entirely different sets of clusters, so the approximations do not have to be related in any way. The rest of this paper will not refer to approximated equivalence relations anymore. Instead, it will simply consider clusters, i.e. sets of items that are considered equivalent by a given clustering, and it will use $\match$ to denote the ideal equivalence relation.

\section{ABCDE inputs}

ABCDE uses three artifacts as inputs:
\begin{enumerate}
    \item A baseline clustering $\Base$, which partitions a set of items into clusters.
    \item An experiment clustering $\Exp$. \\
    The rest of this paper assumes that $\Base$ and $\Exp$ partition the same set of items $T$ into clusters. If this requirement is not met, then preprocessing can restrict the clusterings to the items they have in common, as was done in~\cite{vanstaden2024pointwise}. The preprocessing can report the number (and weight - see the next point) of removed/added items and take samples for debugging.
    \item Application-specific auxiliary information: a mapping that associates each item in $T$ with a weight -- a positive real number that records its importance.
\end{enumerate}

We denote the cluster that contains item $i \in T$ in $\Base$ with $\Base(i)$, and the cluster that contains $i$ in $\Exp$ with $\Exp(i)$. A cluster is simply a set of items that are considered equivalent by the clustering. Hence it is always the case that $i \in \Base(i)$ and $i \in \Exp(i)$.

We denote the weight of an item $i \in T$ with $\weight(i)$, and the weight of a set of items $I \subseteq T$ with $\weight(I) = \sum_{i \in I} \weight(i)$. So $\weight(T)$ is the total weight of all the items of interest, while $\weight(\Base(i))$ is the weight of the cluster in $\Base$ that contains the item $i$.

\subsection{Item weights}

ABCDE does not dictate how weights for the individual items should be obtained. That part is application-specific. However, since the weights are central to all the metrics, this section briefly describes some ideas for weight assignment schemes in practical applications.

In many applications, an item can be associated with an intrinsic importance value. For example, one might consider all items to be equally important, or the importance can be a function of the item's properties, such as its provenance, the kinds and completeness of its data, etc.

In some applications, the clusters that were produced in the past have associated importance/popularity information. The past cluster's importance can then be divided among its members, perhaps proportionally to their intrinsic importance values. To adapt that past weight mapping to a present clustering, we can compute a weight for each present cluster by summing its members' past weights (use zero for new members that don't have a past weight), and then dividing it among all the members to get current per-item weights. Doing that has two benefits: 1) items that didn't exist in the past clustering can get reasonable weights, and 2) items that were unimportant in the past but that now enter important clusters will get boosted weights, which helps to make merges in important clusters prominent. Such a scheme can be used to compute two weights for each item (one for $\Base$ and one for $\Exp$) which can be combined into one final weight for each item, e.g. by taking the maximum of the two weight values.

\section{ABCDE impact metrics}

ABCDE's impact metrics describe the relationship between $\Base$ and $\Exp$ without reference to quality. As will become clear in a moment, the $\JaccardDistance$ characterizes the distance between the two clusterings in a single number, while the $\SplitRate$ and $\MergeRate$ metrics characterize the magnitude of the splits and merges respectively. These metrics do not require human judgements; they can be calculated automatically and exactly.

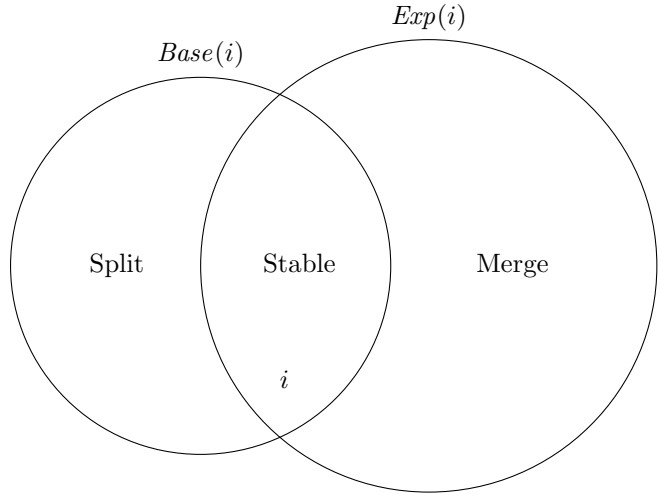
\begin{figure*}[t!]
\centering
\begin{tikzpicture}[fill=gray]
\draw (-1.5,0) circle (2.5)
      (-1.5,2.5)  node [text=black,above] {$\Base(i)$}
      (1.5,0) circle (3)
      (1.5,3)  node [text=black,above] {$\Exp(i)$}
      (-0.4,-1.5) node [text=black] {$i$}
      (-0.2,0.035) node [text=black] {Stable}
      (-2.6,0) node [text=black] {Split}
      (2.6,0) node [text=black] {Merge}
      ;
\end{tikzpicture}
\caption{The clustering situation from the perspective of item $i$. The item $i$ is always in the intersection of $\Base(i)$ and $\Exp(i)$, which is never empty. $\text{Split} = \Base(i) \setminus \Exp(i)$ denotes the set of items that got split off in $\Exp$ from the perspective of $i$. $\text{Merge} = \Exp(i) \setminus \Base(i)$ denotes the set of items that are merged in $\Exp$ from the perspective of $i$. $\text{Stable} = \Base(i) \cap \Exp(i)$ denotes the items that remained stable from the perspective of $i$.}\label{base_exp_cluster_diagram}
\end{figure*}

The pointwise approach~\cite{vanstaden2024pointwise} considers the clustering situation from the vantage point of each item. An item is a member of a $\Base$ cluster and a member of an $\Exp$ cluster, and from its perspective some items are split off, merged, or stable, as depicted in Figure~\ref{base_exp_cluster_diagram}. That classification leads to the definitions of the impact metrics of an item $i$:

\begin{align}
    \SplitRate(i) &= \frac{\weight(\Base(i) \setminus \Exp(i))}{\weight(\Base(i))} \\
    \MergeRate(i) &= \frac{\weight(\Exp(i) \setminus \Base(i))}{\weight(\Exp(i))} \\
    \JaccardDistance(i) &= \frac{\weight(\Base(i) \setminus \Exp(i)) + \weight(\Exp(i) \setminus \Base(i))}{\weight(\Base(i) \cup \Exp(i))}
\end{align}

Informally:

\begin{itemize}
\item $\SplitRate(i)$ is the weight fraction of $\Base(i)$ that got split off in $\Exp$ from the perspective of $i$.
\item $\MergeRate(i)$ is the weight fraction of $\Exp(i)$ that is newly merged in from the perspective of $i$.
\item $\JaccardDistance(i)$ expresses the weight of the split and merged items as a fraction of the split and merged and stable items.
\end{itemize}

Notice that all three of these metrics will be zero when item $i$ did not experience any change, i.e. when $\Base(i) = \Exp(i)$.

In the pointwise clustering metrics, the per-item definitions are lifted to arbitrary sets of items by reporting expected values, i.e. weighted averages. For any set of items $I \subseteq T$, we have:

\begin{align}
    \SplitRate(I) &= \frac{\sum_{i \in I} \mathit{weight}(i) \cdot \SplitRate(i)}{\mathit{weight}(I)} \\
    \MergeRate(I) &= \frac{\sum_{i \in I} \mathit{weight}(i) \cdot \MergeRate(i)}{\mathit{weight}(I)} \\
    \JaccardDistance(I) &= \frac{\sum_{i \in I} \mathit{weight}(i) \cdot \JaccardDistance(i)}{\mathit{weight}(I)}
\end{align}

So $\SplitRate(I)$ is the weighted average $\SplitRate$ of the items in $I$, i.e. the expected $\SplitRate$ of an item in $I$. Taking $I = T$, we obtain an overall $\SplitRate$ metric. We can also obtain $\SplitRate$ metrics for particular slices of items. And we can obtain $\SplitRate$ metrics for individual clusters. The latter can be useful, for example, to see which $\Base$ clusters have the highest $\SplitRate$, or which $\Exp$ clusters have the highest $\MergeRate$.

The simplicity of the pointwise definitions makes the impact metrics easy to interpret. The metrics are also mathematically well-behaved. As mentioned in~\cite{vanstaden2024pointwise}, only  the  relative  magnitudes  of  the  weights  matter for the metrics, and the metrics of aggregates compose nicely because they are based on expected values. The definitions of $\SplitRate$ and $\MergeRate$ are perfectly symmetric: the $\SplitRate$ is the same as the $\MergeRate$ when the roles of the $\Base$ and $\Exp$ clusterings are swapped. The $\JaccardDistance$ is a true distance metric on the set of all clusterings of a fixed set of weighted items, as was proved in~\cite{vanstaden2024pointwise}.

The $\SplitRate$ and $\MergeRate$ metrics are closely related to the $\Precision$ and $\Recall$ metrics of~\cite{vanstaden2024pointwise}. To see that, pretend for a moment that $\Base$ is a ground truth clustering, and we want to evaluate $\Exp$ with respect to it.  Then the pointwise $\OverMergeRate$ of~\cite{vanstaden2024pointwise}, which is the same as $1 - \Precision$, is exactly the $\MergeRate$ of this paper, and the $\UnderMergeRate$, which is the same as $1 - \Recall$, coincides exactly with the $\SplitRate$ of this paper. So ABCDE's impact metrics don't reinvent the wheel. Instead, they embrace established metrics and adapt them to characterize aspects of a clustering change that are interesting and meaningful for developers.

\subsection{Insights and debugging}

The impact evaluation can generate a report that clearly shows the main impact metrics, namely the overall $\SplitRate$, $\MergeRate$ and $\JaccardDistance$. It can also show the most affected clusters of each clustering. For example, it can present information about the 100 clusters of $\Base$ that contribute the most to the overall $\JaccardDistance$, where the contribution of a cluster $C$ is given by $\frac{\weight(C) \cdot \JaccardDistance(C)}{\weight(T)}$, and similarly for $\Exp$\footnote{Another possibility is to show the most affected clusters of $\Base$ and $\Exp$ in a single table that ranks them according to their contributions. If a large and important cluster of $\Base$ is split up, then it will typically show up in the most affected clusters of $\Base$. Likewise, a heavy cluster that is created by merging together many lighter clusters will typically show up in the most affected clusters of $\Exp$. So the most affected clusters of $\Base$ and $\Exp$ complement each other well and they can interleaved in the report.}. It can also surface metrics for important slices of items when such slices are known in advance. And of course it can provide the ability to look up the impact metrics of individual items.

\subsubsection{Exploring the impact}
In many practical applications, the items that are clustered can each be associated with simple attributes. For example, when clustering images, each image can be associated with a size (small, medium, large), a list of its dominant colors, its provenance, its dimensions (width and height in pixels), etc. These attributes can be used to define slices of items. An example slice would be the large images from Wikipedia that contain yellow as a dominant color and whose width do not exceed 512 pixels. For that slice of images, one might be interested to know the average height of the image in pixels, or a histogram of the other dominant colors, etc. Even a small set of simple attributes can define a vast number of slices, and similarly there are many possibilities for what to observe about the slice. To keep data exploration practical, one can use a sample of the items (or the whole population if it is small enough) and interactively specify the criteria to slice it and what to observe about the slice.

The fact that the individual items have weights and ABCDE impact metrics fits in nicely with this approach to data exploration. A human can interactively slice and dice (a sample of) the items to understand how various sets of items are impacted by the clustering change. For example, one can see the $\weight$, the $\SplitRate$, $\MergeRate$ and $\JaccardDistance$ of the large images from Wikipedia that contain yellow as a dominant color and whose width do not exceed 512 pixels.
Instead of starting from the intrinsic attributes of items and looking at the impact metrics, one can also start from the impact metrics and explore which kinds of items were impacted and how. This kind of exploration can be very useful in practice, especially when the population of items is huge, diverse, and the effect of a clustering change is not obvious. The rest of this section gives a bit more detail about how to facilitate such explorations.

The whole population of items $T$ can be partitioned into two sub-populations, namely those affected by the clustering change, and those that remain unaffected:
\begin{align}
\mathit{AffectedItems} &= \{i \in T | \Base(i) \neq \Exp(i)\} \\
\mathit{UnaffectedItems} &= \{i \in T | \Base(i) = \Exp(i)\}
\end{align}

In practical applications, there can be billions of items, and a clustering change will typically affect only a relatively small fraction of them. 
Naively sampling items from $T$ according to their weight will therefore result in a sample in which most items are unaffected by the change. 
The naive sampling can be refined to sample only from the $\mathit{AffectedItems}$, but the resulting observations can still have a high variance when many items experienced only small impact.

To overcome that, one can refine the technique further and use \textit{importance sampling}: instead of sampling an affected item $i$ at each draw with a probability equal to 
\begin{align}
p(i) = \frac{\weight(i)}{\sum_{j \in \mathit{AffectedItems}} \weight(j)}
\end{align}
one can sample it with a probability equal to
\begin{align}
q(i) = \frac{\weight(i) \cdot \JaccardDistance(i)}{\sum_{j \in \mathit{AffectedItems}} \weight(j) \cdot \JaccardDistance(j)}
\end{align}
That takes into account not only the weight of the item, but also how much it was impacted by the clustering change. For all $i \in \mathit{AffectedItems}$, we have $q(i) > 0$, so sampling according to $q$ won't exclude any affected items, which is necessary for correct results. Let $m$ denote an arbitrary ABCDE impact metric, i.e. $\SplitRate$, $\MergeRate$, or $\JaccardDistance$. Notice that
\begin{align}
m(\mathit{AffectedItems}) = \sum_{i \in \mathit{AffectedItems}} p(i) \cdot m(i)
\end{align}
The importance sampling is based on the observation that
\begin{align}
m(\mathit{AffectedItems}) = \sum_{i \in \mathit{AffectedItems}} q(i) \cdot m(i) \cdot (p(i) / q(i))
\end{align}
Importance sampling says that, instead of sampling affected items according to $p(i)$ and computing the average of $m(i)$ over the sample to obtain an estimate $\hat{m}(\mathit{AffectedItems})$, we can sample according to $q(i)$ and compute the average of $m(i) \cdot (p(i) / q(i))$ over the sample.

We ultimately want to estimate the impact metrics of the whole population of items, i.e. we are interested obtaining the estimates $\hat{m}(T)$. That is fortunately simple.
All three of the ABCDE impact metrics are zero for unaffected items:
\begin{align}
i \in \mathit{UnaffectedItems} \Longrightarrow m(i) = 0
\end{align}
and hence
\begin{align}
m(\mathit{UnaffectedItems}) = 0
\end{align}
The impact metrics are lifted to sets of items by using expected values (weighted averages), which compose nicely and make it simple to obtain a metric for the whole population of items from the metric of the affected sub-population:
\begin{align}
m(T) &= \frac{\weight(\mathit{AffectedItems}) \cdot m(\mathit{AffectedItems}) + \weight(\mathit{UnaffectedItems}) \cdot m(\mathit{UnaffectedItems})}{\weight(\mathit{AffectedItems}) + \weight(\mathit{UnaffectedItems})} \\
&= \frac{\weight(\mathit{AffectedItems}) \cdot m(\mathit{AffectedItems}) + 0}{\weight(T)} \\
&= m(\mathit{AffectedItems}) \frac{\weight(\mathit{AffectedItems})}{\weight(T)}
\end{align}

Using that with the equation from before, we have
\begin{align}
m(T) &= \left(\sum_{i \in \mathit{AffectedItems}} q(i) \cdot m(i) \cdot (p(i) / q(i))\right) \frac{\weight(\mathit{AffectedItems})}{\weight(T)} \\
&= \sum_{i \in \mathit{AffectedItems}} q(i) \cdot m(i) \cdot \left(\frac{p(i)}{q(i)} \frac{\weight(\mathit{AffectedItems})}{\weight(T)} \right)
\end{align}
So importance sampling says that we can sample items\footnote{Unaffected items will have $q(i) = 0$, so sampling from all items according to $q$ is the same as sampling from the affected items according to $q$.} according to $q(i)$ and compute the average over the sample of
$$m(i) \cdot \frac{p(i)}{q(i)} \frac{\weight(\mathit{AffectedItems})}{\weight(T)}$$
in order to obtain an estimate $\hat{m}(T)$ for the whole population.

Simplifying the first fraction of that expression yields:
\begin{align}
\frac{p(i)}{q(i)} &= \frac{\weight(i)}{\sum_{j \in \mathit{AffectedItems}} \weight(j)} \frac{\sum_{j \in \mathit{AffectedItems}} \weight(j) \cdot \JaccardDistance(j)}{\weight(i) \cdot \JaccardDistance(i)} \\
&= \frac{\weight(i)}{\weight(i) \cdot \JaccardDistance(i)} \frac{\sum_{j \in \mathit{AffectedItems}} \weight(j) \cdot \JaccardDistance(j)}{\sum_{j \in \mathit{AffectedItems}} \weight(j)} \\
&= \frac{1}{\JaccardDistance(i)} \JaccardDistance(\mathit{AffectedItems})
\end{align}
which helps to simplify the whole expression:
\begin{align}
& m(i) \cdot \frac{p(i)}{q(i)} \frac{\weight(\mathit{AffectedItems})}{\weight(T)} \nonumber \\
&= m(i) \cdot \frac{1}{\JaccardDistance(i)} \JaccardDistance(\mathit{AffectedItems}) \frac{\weight(\mathit{AffectedItems})}{\weight(T)} \\
&= m(i) \cdot \frac{\JaccardDistance(T)}{\JaccardDistance(i)}
\end{align}

Sampling with replacement is useful for practical applications when the items and/or the clusters can have a broad range of weights. Instead of working with the sampled items before deduplication and using a simple average for the estimate, we can work with deduplicated items, i.e. unique sampled items, where each one has a draw count $dc(i) \in \mathbb{N}$. Then we have to use the $dc$-weighted average for the estimate, which is the same as having an additional multiplier $\frac{dc(i)}{\sum_{j \in \mathit{SampledItems}} dc(j)}$ for the metric of each unique sampled item and using the sum for the estimate.

Summary of the sampling: We sample items with replacement. At each round, the item $i$ is drawn with a probability of $q(i)$. After sampling, we have a set of unique sampled items $\mathit{SampledItems}$ where each one is associated with a draw count $dc(i)$. The size of the sample can be fairly large, say a million unique items or so, the main constraint on its size being that is should facilitate interactive slicing. For each $i \in \mathit{SampledItems}$, we note down its importance weight
\begin{align}
w(i) = \frac{dc(i)}{\sum_{j \in \mathit{SampledItems}} dc(j)} \frac{\JaccardDistance(T)}{\JaccardDistance(i)}
\end{align}
and its ABCDE impact metrics.

Each unique sampled item $i$ contributes a value of $w(i) \cdot m(i)$ to the overall estimate $\hat{m}(T)$, i.e. $\hat{m}(T) = \sum_{i \in \mathit{SampledItems}} w(i) \cdot m(i)$. When a user interactively specifies constraints for a slice, then it is fast to restrict the sample to get $\mathit{Slice} \subseteq \mathit{SampledItems}$ and report $\sum_{i \in \mathit{Slice}} w(i) \cdot m(i)$. Likewise, it is fast to group the $\mathit{SampledItems}$ or $\mathit{Slice}$ by attribute values and to report the value of $\sum_{i \in \mathit{Group}} w(i) \cdot m(i)$ for each group.
Displaying the top 10 or 20 groups with the largest values can be very useful for understanding the breakdown of the impact of the clustering change. That can also provide the user with ideas for drilling deeper into the impact. Moreover, each slice or group is backed by a set of sampled items, i.e. concrete examples, that can help in debugging. These items can be subsampled with a selection probability proportional to $w(i)$ if needed.


\section{ABCDE quality metrics}

The previous section characterized the clustering diff with impact metrics. The impact metrics characterize the magnitude of the clustering change: they say how big or small the overall $\SplitRate$ and the $\MergeRate$ are, for example.

This section characterizes the \textit{quality} of the clustering diff: the goal is to find out whether the changes to the clustering are good or bad. To do that, ABCDE will sample pairs of items, and a human can consider the two items of a pair and judge whether or not they should be considered equivalent (i.e. whether the item pair is in the underlying equivalence relation or not). On the basis of the human judgements, ABCDE will report quality metrics. For example, it will decompose the overall $\SplitRate$ into two components, a $\GoodSplitRate$ and a $\BadSplitRate$, such that
$$\SplitRate(T) = \GoodSplitRate(T) + \BadSplitRate(T)$$
and similarly it will decompose the $\MergeRate$ into a $\GoodMergeRate$ and a $\BadMergeRate$.

The focus is mostly on estimating overall quality metrics, and not so much on interactive exploration of the quality. The reason is that, because of limited time and cost, humans can judge the equivalence of only a very limited number of item pairs, say several hundreds or a few thousands. So there will typically be too few judgements about a particular slice to report quality metrics for it with high confidence. Conceptually, the human judgements provide information about the ideal clustering over all items, i.e. a giant ground truth clustering which does not exist in a materialized form. The item pairs that are judged are chosen carefully on the basis of the clustering diffs, as we will see. 

In applications where $\Precision$, i.e. cluster homogeneity, is not very important, the $\GoodSplitRate$, $\BadSplitRate$, $\GoodMergeRate$ and $\BadMergeRate$ metrics, which we will abbreviate henceforth as $\GoodBadSplitMergeRate$, might be enough to judge the quality of the clustering change. However, many practical applications need to maintain a high clustering $\Precision$ -- an $\Exp$ that lowers it significantly is essentially worthless. $\Precision$ losses are typically very hard to detect with ground truth clusterings, because the items of a ground truth clustering are selected in advance: it is hard/impossible to foresee the kinds of overmerges that experiments will perform later on, and the items that participate in overmerges were perhaps not even part of the population at the time when the ground truth clustering was created\footnote{$\Recall$ losses are much easier to measure with ground truth clusterings, because a ground truth cluster captures a set of items that ought to stick together, and $\Recall$ will drop if $\Exp$ splits the set up. So as long as a ground truth clustering includes clusters that are important, it will monitor the $\Recall$ of important things. $\Precision$ losses are different, because it is hard/impossible to know in advance which items will be wrongly added to important clusters in the future.}. This is where ABCDE shines: it can estimate the difference in $\Precision$ between the two clusterings
\begin{align}
\DeltaPrecision(T) = \Precision_{\Exp}(T) - \Precision_{\Base}(T)
\end{align}
Naively, $\DeltaPrecision(T)$ can be computed by estimating $\Precision_{\Exp}(T)$ and $\Precision_{\Base}(T)$ separately and subtracting the estimates. But that yields a very noisy result, because the vast majority of human judgements will be spent on items that are unaffected by the clustering change at hand. ABCDE uses a more sophisticated approach where the sampling of item pairs for human judgement is informed by the diffs between $\Base$ and $\Exp$. The technique harmonizes well with the other ABCDE quality metrics: ABCDE will sample a set of item pairs to estimate $\DeltaPrecision(T)$, and subsets of that sample are used to estimate the $\GoodBadSplitMergeRate$. Of course it is also possible to ask humans to judge only the sampled pairs that are needed to estimate the $\GoodBadSplitMergeRate$ if only that is needed.

The rest of this section uses some additional notation:
\begin{itemize}
    \item $\indicate(e)$ is the indicator function for the expression $e$; it is defined to be 1 if $e$ is true, 0 otherwise.
    \item $i \match j$ is true if and only if items $i$ and $j$ are equivalent according to the underlying equivalence relation, while $i \distinct j$ is true if an only if items $i$ and $j$ are not equivalent.
\end{itemize}

The rest of this section fleshes out the technical details of the sampling and estimation approach, and mentions some practical considerations.

\subsection{Estimating a weighted sum from a weighted sample}\label{EstimatingWeightedSum}

Many of the formal manipulations below use a sampling approach to estimate a metric that is expressed as a weighted sum. On a high level, the estimation procedure has three steps:
\begin{itemize}
\item Take a weighted sample of the population.
\item Compute the average (mean) over the sample.
\item Multiply the average by the total weight of the population to get an estimate of the metric.
\end{itemize}

The reasoning behind the procedure is as follows.  Suppose we want to estimate the sum:
$$\sum_{x} w_x f(x)$$
To do that we can take a weighted sample of elements $x_i$, which would let us then compute the mean:
$$\hat{f}(x) = \frac{\sum_{x_i} f(x_i)}{N}$$
To get an estimate of the metric we originally cared about, we can just see that:
$$E[\hat{f}(x)] = \frac{\sum_{x} w_x f(x)}{\sum_{x} w_x }$$
$$\sum_{x} w_x f(x) = E[\hat{f}(x)] \sum_{x} w_x$$
So the original sum can be estimated by the mean over a weighted sample, multiplied by the total weight of the whole population of elements that we were sampling over.

\subsection{The $\Precision$ of a single clustering}

For a given clustering $C$, we can compute the absolute precision of an item $i$ as follows:
\begin{align}
\Precision(i) = \sum_{j \in C(i)} \frac{\weight(j)}{\weight(C(i))} \indicate\left(i \match j\right)
\end{align}
There is nothing new here: it is the same as the $\Precision(i)$ defined in~\cite{vanstaden2024pointwise}, but written in a form that uses item pairs and the underlying equivalence relation.

Overall precision is just the weighted average of the $\Precision$ over all the items:
\begin{align}
\Precision(T) = \sum_{i \in T} \frac{\weight(i)}{\weight(T)} \sum_{j \in C(i)} \frac{\weight(j)}{\weight(C(i))} \indicate\left(i \match j\right)
\end{align}

\subsection{$\DeltaPrecision$ between two clusterings}

Given an $\Exp$ and a $\Base$ clustering, the difference in precision for an item $i$ is given by:
\begin{align}
& \DeltaPrecision(i) \nonumber \\
= & \sum_{j \in \Exp(i)} \left( \frac{\weight(j)}{\weight(\Exp(i))} \indicate\left(i \match j\right) \right) - \sum_{j \in \Base(i)} \left( \frac{\weight(j)}{\weight(\Base(i))} \indicate\left(i \match j\right) \right) \\
= & \sum_{j \in \Exp(i) \setminus \Base(i)} \left( \frac{\weight(j)}{\weight(\Exp(i))} \indicate\left(i \match j\right) \right)
- \sum_{j \in \Base(i) \setminus \Exp(i)} \left( \frac{\weight(j)}{\weight(\Base(i))} \indicate\left(i \match j\right) \right)
\nonumber \\ &
\ + \sum_{j \in \Base(i) \cap \Exp(i)} \left( \left(\frac{\weight(j)}{\weight(\Exp(i))} - \frac{\weight(j)}{\weight(\Base(i))}\right) \indicate\left(i \match j\right)\right)
\end{align}
The last term can be simplified a bit:
\begin{align*}
\frac{\weight(j)}{\weight(\Exp(i))} - \frac{\weight(j)}{\weight(\Base(i))} = \frac{\weight(\Base(i)) -  \weight(Exp(i))}{\weight(\Base(i)) \cdot \weight(Exp(i))} \weight(j)
\end{align*}
We can write $\DeltaPrecision(i)$ as a standard weighted sum by introducing two new variables, a weight $u_j$ and label $l_j$.  Let:
\begin{align}
\mathrm{if} \ j \in \Exp(i) \setminus \Base(i)\ :&\ u_{j} = \frac{\weight(j)}{\weight(\Exp(i))},\ l_{j} = 1  \\
\mathrm{if} \ j \in \Base(i) \setminus \Exp(i)\ :&\ u_{j} = \frac{\weight(j)}{\weight(\Base(i))},\ l_{j} = -1  \\
\mathrm{if} \ j \in \Base(i) \cap \Exp(i)\ :&\ u_{j} = \frac{|\weight(\Base(i)) -  \weight(Exp(i))|}{\weight(\Base(i)) \cdot \weight(Exp(i))} \weight(j) \nonumber \\
&\ l_{j} = \mathrm{sgn}(\weight(\Base(i)) -  \weight(Exp(i)))
\end{align}
Then
\begin{align}
\DeltaPrecision(i) = \sum_{j \in \Base(i) \cup \Exp(i)} u_{j} l_{j} \indicate\left(i \match j\right)
\end{align}

The overall $\DeltaPrecision$ is the weighted average $\DeltaPrecision$ of the individual items:
\begin{align}
\DeltaPrecision(T) = \sum_{i \in T} \frac{\weight(i)}{\weight(T)} \DeltaPrecision(i)
\end{align}
Given a pair of items $i \in T$ and $j \in \Base(i) \cup \Exp(i)$, we can define a weight and a label for the pair:
\begin{align}
\mathrm{if} \ j \in \Exp(i) \setminus \Base(i)\ :&\ u_{ij} = \frac{\weight(i)}{\weight(T)} \frac{\weight(j)}{\weight(\Exp(i))},\ l_{ij} = 1  \\
\mathrm{if} \ j \in \Base(i) \setminus \Exp(i)\ :&\ u_{ij} = \frac{\weight(i)}{\weight(T)} \frac{\weight(j)}{\weight(\Base(i))},\ l_{ij} = -1  \\
\mathrm{if} \ j \in \Base(i) \cap \Exp(i)\ :&\ u_{ij} = \frac{\weight(i)}{\weight(T)} \frac{|\weight(\Base(i)) -  \weight(Exp(i))|}{\weight(\Base(i)) \cdot \weight(Exp(i))} \weight(j) \nonumber \\
&\ l_{ij} = \mathrm{sgn}(\weight(\Base(i)) -  \weight(Exp(i)))
\end{align}
Overall $\DeltaPrecision$ can then be written as a weighted sum:
\begin{align}
\DeltaPrecision(T) = \sum_{i \in T} \sum_{j \in \Base(i) \cup \Exp(i)} u_{ij} l_{ij} \indicate(i \match j)
\end{align}

So we can sample item pairs $i, j$ according to $u_{ij}$ and estimate $\DeltaPrecision(T)$ by computing the average of
$l_{ij} \indicate\left(i \match j\right)$ over the sample and multiplying it by 
$$\sum_{i \in T} \sum_{j \in \Base(i) \cup \Exp(i)} u_{ij}$$
(See the explanation in section~\ref{EstimatingWeightedSum} above for the justification.)

\subsection{$\GoodBadSplitMergeRate$ metrics of individual items}

Recall from the section on impact metrics that
$$\SplitRate(i) = \frac{\weight(\Base(i) \setminus \Exp(i))}{\weight(\Base(i))}$$
We can also express the split rate in terms of pairs of items:
\begin{align}
\SplitRate(i) = \sum_{j \in \Base(i) \setminus \Exp(i)} \frac{\weight(j)}{\weight(\Base(i))}
\end{align}

The $\GoodSplitRate$ of an item $i$ represents the weight fraction of the base cluster that was correctly split off from $i$'s point of view:
\begin{align}
\GoodSplitRate(i) = \sum_{j \in \Base(i) \setminus \Exp(i)} \frac{\weight(j)}{\weight(\Base(i))} \indicate\left(i \distinct j\right)
\end{align}

We can similarly define the $\BadSplitRate$, and similar rates for merges:
\begin{align}
\BadSplitRate(i) &= \sum_{j \in \Base(i) \setminus \Exp(i)} \frac{\weight(j)}{\weight(\Base(i))} \indicate\left(i \match j\right) \\
\GoodMergeRate(i) &= \sum_{j \in \Exp(i) \setminus \Base(i)} \frac{\weight(j)}{\weight(\Exp(i))} \indicate\left(i \match j\right) \\
\BadMergeRate(i) &= \sum_{j \in \Exp(i) \setminus \Base(i)} \frac{\weight(j)}{\weight(\Exp(i))} \indicate\left(i \distinct j\right)
\end{align}

They have a straightforward relationship with the $\SplitRate$ and $\MergeRate$ impact metrics:
\begin{align}
\SplitRate(i) &= \GoodSplitRate(i) + \BadSplitRate(i) \\
\MergeRate(i) &= \GoodMergeRate(i) + \BadMergeRate(i)
\end{align}

\subsection{Lifted $\GoodBadSplitMergeRate$ metrics}

The per-item $\GoodBadSplitMergeRate$ metrics are pointwise metrics that are lifted to arbitrary sets of items via expected values. For example, the overall $\GoodSplitRate$ metric is simply the weighted average of the per-item $\GoodSplitRate$ metrics:
\begin{align}
\GoodSplitRate(T) &= \sum_{i \in T} \frac{\weight(i)}{\weight(T)} \GoodSplitRate(i) \\
&= \sum_{i \in T} \frac{\weight(i)}{\weight(T)} \sum_{j \in \Base(i) \setminus \Exp(i)} \frac{\weight(j)}{\weight(\Base(i))} \indicate\left(i \distinct j\right) \\
&= \sum_{i \in T} \sum_{j \in \Base(i) \setminus \Exp(i)} \frac{\weight(i) \cdot \weight(j)}{\weight(T) \cdot \weight(\Base(i))} \indicate\left(i \distinct j\right)
\end{align}

The equations that held for individual items also lift to arbitrary sets of items. For example:
\begin{align}
\SplitRate(T) &= \GoodSplitRate(T) + \BadSplitRate(T) \\
\MergeRate(T) &= \GoodMergeRate(T) + \BadMergeRate(T)
\end{align}

\subsection{Estimating $\GoodBadSplitMergeRate$ metrics}

\subsubsection{Split metrics}

We focus on estimating $\GoodSplitRate(T)$. Estimating $\BadSplitRate(T)$ is very similar, or we can simply use $\SplitRate(T) - \GoodSplitRate(T)$. Recall that:

$$\GoodSplitRate(T) = \sum_{i \in T} \sum_{j \in \Base(i) \setminus \Exp(i)} \frac{\weight(i) \cdot \weight(j)}{\weight(T) \cdot \weight(\Base(i))} \indicate\left(i \distinct j\right)$$

So to obtain an estimate of $\GoodSplitRate(T)$, we can sample pairs of items $i \in T$ and $j \in \Base(i) \setminus \Exp(i)$, where the weight of a pair $(i, j)$ is given by
$$\frac{\weight(i) \cdot \weight(j)}{\weight(T) \cdot \weight(\Base(i))}$$
and proceed by computing the average of $\indicate\left(i \distinct j\right)$ over the sample and multiplying it by
\begin{align*}
& \sum_{i \in T} \sum_{j \in \Base(i) \setminus \Exp(i)} \frac{\weight(i) \cdot \weight(j)}{\weight(T) \cdot \weight(\Base(i))} \\
&= \sum_{i \in T} \frac{\weight(i)}{\weight(T)} \sum_{j \in \Base(i) \setminus \Exp(i)} \frac{\weight(j)}{\weight(\Base(i))} \\
&= \sum_{i \in T} \frac{\weight(i)}{\weight(T)} \SplitRate(i) \\
&= \SplitRate(T)
\end{align*}

In the sampling strategy for estimating $\DeltaPrecision(T)$ from before, we had:
$$
\mathrm{if} \ j \in \Base(i) \setminus \Exp(i)\ :\ u_{ij} = \frac{\weight(i)}{\weight(T)} \frac{\weight(j)}{\weight(\Base(i))},\ l_{ij} = -1
$$
So we can use the human judgements of these sampled pairs (and ignore the label value $l_{ij} = -1$) to estimate $\GoodSplitRate(T)$, since the sampling weights $u_{ij}$ are exactly what we want.

\subsubsection{Merge metrics}

Merges are analogous to splits; we include this section purely for completeness.

We focus on estimating $\GoodMergeRate(T)$. Estimating $\BadMergeRate(T)$ is very similar, or we can simply use $\MergeRate(T) - \GoodMergeRate(T)$. We have:

$$\GoodMergeRate(T) = \sum_{i \in T} \sum_{j \in \Exp(i) \setminus \Base(i)} \frac{\weight(i) \cdot \weight(j)}{\weight(T) \cdot \weight(\Exp(i))} \indicate\left(i \match j\right)$$

So to obtain an estimate of $\GoodMergeRate(T)$, we can sample pairs of items $i \in T$ and $j \in \Exp(i) \setminus \Base(i)$, where the weight of a pair $(i, j)$ is given by
$$\frac{\weight(i) \cdot \weight(j)}{\weight(T) \cdot \weight(\Exp(i))}$$
and proceed by computing the average of $\indicate\left(i \match j\right)$ over the sample and multiplying it by
\begin{align*}
& \sum_{i \in T} \sum_{j \in \Exp(i) \setminus \Base(i)} \frac{\weight(i) \cdot \weight(j)}{\weight(T) \cdot \weight(\Exp(i))} \\
&= \sum_{i \in T} \frac{\weight(i)}{\weight(T)} \sum_{j \in \Exp(i) \setminus \Base(i)} \frac{\weight(j)}{\weight(\Exp(i))} \\
&= \sum_{i \in T} \frac{\weight(i)}{\weight(T)} \MergeRate(i) \\
&= \MergeRate(T)
\end{align*}

In the sampling strategy for estimating $\DeltaPrecision(T)$ from before, we had:
$$
\mathrm{if} \ j \in \Exp(i) \setminus \Base(i)\ :\ u_{ij} = \frac{\weight(i)}{\weight(T)} \frac{\weight(j)}{\weight(\Exp(i))},\ l_{ij} = 1
$$
So we can use the human judgements of these sampled pairs (and ignore the label value $l_{ij} = 1$) to estimate $\GoodMergeRate(T)$, since the sampling weights $u_{ij}$ are exactly what we want.

\subsection{Summary of the estimation of quality metrics}

We define a population of item pairs for a clustering change as follows:
\begin{align}
\mathrm{AllPairs} = \{(i, j) | i \in T \land j \in \Base(i) \cup \Exp(i)\}
\end{align}
It can be partitioned into three sub-populations:
\begin{align}
\mathrm{SplitPairs} &= \{(i, j) | i \in T \land j \in \Base(i) \setminus \Exp(i)\} \\
\mathrm{MergePairs} &= \{(i, j) | i \in T \land j \in \Exp(i) \setminus \Base(i)\} \\
\mathrm{StablePairs} &= \{(i, j) | i \in T \land j \in \Base(i) \cap \Exp(i)\}
\end{align}

For each pair $(i, j) \in \mathrm{AllPairs}$, compute the weight $u_{ij}$ and the label $l_{ij}$ as follows:

\begin{align}
\mathrm{if} \ (i,j) \in \mathrm{SplitPairs}\ :&\ u_{ij} = \frac{\weight(i)}{\weight(T)} \frac{\weight(j)}{\weight(\Base(i))},\ l_{ij} = -1  \\
\mathrm{if} \ (i,j) \in \mathrm{MergePairs}\ :&\ u_{ij} = \frac{\weight(i)}{\weight(T)} \frac{\weight(j)}{\weight(\Exp(i))},\ l_{ij} = 1  \\
\mathrm{if} \ (i,j) \in \mathrm{StablePairs}\ :&\ u_{ij} = \frac{\weight(i)}{\weight(T)} \frac{|\weight(\Base(i)) -  \weight(Exp(i))|}{\weight(\Base(i)) \cdot \weight(Exp(i))} \weight(j) \nonumber \\
&\ l_{ij} = \mathrm{sgn}(\weight(\Base(i)) -  \weight(Exp(i)))
\end{align}

Sample pairs $(i, j)$ according to their weight $u_{ij}$ to obtain a multiset\footnote{Sampling with replacement is recommended and discussed later.} of sampled pairs $\mathrm{SampledPairs}$. 

After humans judged the equivalence of the sampled pairs, we can estimate the various overall quality metrics as follows:

\begin{itemize}
\item  For $\DeltaPrecision(T)$, we compute the average of
   $ l_{ij} \indicate\left(i \match j\right) $
   over all sampled pairs $(i, j)$ and multiply it by
   $$\sum_{(i, j) \in \mathrm{AllPairs}} u_{ij}$$
\item For $\GoodSplitRate(T)$, we compute the average of
   $ \indicate\left(i \distinct j\right) $
   over all sampled pairs $(i, j)$ that are also in $\mathrm{SplitPairs}$ and multiply it by $\SplitRate(T)$.
\item For $\BadSplitRate(T)$, we compute the average of
   $ \indicate\left(i \match j\right) $
   over all sampled pairs $(i, j)$ that are also in $\mathrm{SplitPairs}$ and multiply it by $\SplitRate(T)$.
\item For $\GoodMergeRate(T)$, we compute the average of
   $ \indicate\left(i \match j\right) $
   over all sampled pairs $(i, j)$ that are also in $\mathrm{MergePairs}$ and multiply it by $\MergeRate(T)$.
\item For $\BadMergeRate(T)$, we compute the average of
   $ \indicate\left(i \distinct j\right) $
   over all sampled pairs $(i, j)$ that are also in $\mathrm{MergePairs}$ and multiply it by $\MergeRate(T)$.
\end{itemize}

\subsection{Practical considerations}

\begin{itemize}
\item The $\mathrm{StablePairs}$ include the set $\mathrm{SelfPairs} = \{(i, i) | i \in T\}$, in which each item is paired with itself. So in the definition of $\mathrm{StablePairs}$ above, \textit{please do not naively assume} $i \neq j$.
\item The $\mathrm{SampledPairs}$ typically include many pairs from $\mathrm{SelfPairs}$ in practice. We don't need to get these pairs judged by humans, because $i \match i$ always holds, but it is very important to keep these samples around and to treat them all as if they received a human judgement of $i \match j$.
\item All the pair weight terms $u_{ij}$ divide by the constant $\weight(T)$. We can omit that as long as we remember to divide the multiplier in the $\DeltaPrecision(T)$ computation by $\weight(T)$.
\item Sometimes humans are uncertain and cannot decide whether $i \match j$ or $i \distinct j$. Sometimes it is even impossible to pose the question, for example when the data of $i$ and/or $j$ is not available anymore. In such cases it makes sense to exclude these sampled pairs when estimating the metrics.\\
\textbf{Caveat}: For $\DeltaPrecision(T)$, the remaining sampled pairs can be biased, because all the sampled $\mathrm{SelfPairs}$ will remain (in practice they can easily comprise 30\% of all the sampled pairs and they always get the ``judgement" $i \match j$). This can be remedied by classifying the sampled pairs into classes, e.g. $\mathrm{SelfPairs}$, $\mathrm{SplitPairs}$, $\mathrm{MergePairs}$, $\mathrm{IntersectionPairs}$ (for $\mathrm{StablePairs} \setminus \mathrm{SelfPairs}$), and introducing weights for the remaining sampled pairs such that the total weight of the remaining pairs in each class is equal to the total weight of the originally sampled pairs in each class. For example, if we sampled 1000 split pairs, but only 800 have judgements, then each of the remaining split pairs will get a weight of 1000/800 = 1.25.
\item The technique of weighting the sampled pairs mentioned in the previous point can be applied to arbitrary classes/slices of sampled pairs: if is hard to obtain judgements for some slices, then the weights can ensure that these slices don't get underrepresented in the metrics.
\item It is possible to report confidence intervals by computing the standard errors of the metrics. One very useful result is that $\mathrm{StdErr}(c \cdot X) = c \cdot \mathrm{StdErr}(X)$, which holds because $\mathrm{StdDev}(c \cdot X) = c \cdot \mathrm{StdDev}(X)$. For instance:
  \begin{itemize}
  \item The standard error of $\GoodSplitRate(T)$ is $\SplitRate(T)$ times the standard error of $\indicate\left(i \distinct j\right)$ over all sampled pairs $(i, j)$ that are also $\mathrm{SplitPairs}$.
  \item For $\DeltaPrecision(T)$, we compute the standard error of $ l_{ij} \indicate\left(i \match j\right)$ over all sampled pairs $(i, j)$ and multiply it by $\left(\sum_{(i, j) \in \mathrm{AllPairs}} u_{ij}\right)$.\\ The removal of the sampled pairs without judgements and the subsequent weighting of the remaining pairs means that we must use the standard error of the weighted mean. In particular, we can use the formula from~\cite{standarderrorofweightedmean} for Case I, i.e. the case where the weights indicate the relative importance of the observations:
  \begin{align*}
  \overline{x}_\mathit{wtd} &= \frac{\sum_{i=1}^n w_i x_i}{\sum_{i=1}^n w_i} \\
  \mathrm{StdErr} &= \sqrt{\left(\frac{\sum_{i=1}^n w_i x_i^2}{\sum_{i = 1}^n w_i} - \left(\overline{x}_\mathit{wtd} \right)^2 \right) \frac{\sum_{i=1}^n \left( w_i^2 \right)}{\left(\sum_{i=1}^n w_i \right)^2 - \sum_{i=1}^n \left( w_i^2 \right)}}
  \end{align*}
  \end{itemize}
Note that these confidence intervals quantify only the uncertainty inherent in the sampling. They don't quantify the uncertainty in the human judgements. It is possible to quantify that by replicating questions (i.e. asking multiple humans the same question) and using bootstrapping techniques, but that is impractical unless the budget for human judgements is large.
\item We recommend sampling with replacement. In practical applications the pairs can have a broad range of weights, and it is common to see pairs with a draw count greater than one.
\end{itemize}

\section{Conclusion}

ABCDE is a novel technique for evaluating two whole clusterings -- a Baseline clustering and an Experiment clustering -- that can involve billions of items. Its metrics fully embrace the fact that some items are more important than others in real applications, and that not all clustering diffs/wins/losses are equal. 
ABCDE characterizes the magnitude of the clustering diff with impact metrics that are intuitively meaningful, such as the $\SplitRate$ and the $\MergeRate$. It can also facilitate an interactive exploration of the diff, which is useful to gain more insight into the clustering changes and to debug them. ABCDE can additionally characterize the quality of the clustering diff with metrics that are intuitively meaningful, such as the $\GoodSplitRate$ and the $\BadSplitRate$. It can provide a statistical estimate of $\DeltaPrecision$, a delta quality metric for cluster homogeneity that is important in many practical applications and that is hard to measure accurately with ground truth clusterings. The quality metrics rely on human judgement, but ABCDE rigorously specifies how to sample the pairs of items whose equivalence must be judged, and how to compute quality metrics from the finished judgements. Many of ABCDE's attractive properties stem from the fact that it makes heavy use of pointwise clustering metrics~\cite{vanstaden2024pointwise}. The pointwise metrics are mathematically well-behaved and provide a rigorous foundation for ABCDE.

\bibliographystyle{plain}
\bibliography{main}

\appendix

\section{Sampling at scale}

Sampling at scale is useful for the impact metrics and central to the quality metrics of ABCDE. To explore the impact, we take a sample with replacement of about a million unique items from a population of billions of items. To estimate the quality, we take a sample with replacement of many thousands of unique item pairs from a population of hundreds of billions. Fortunately, these operations are embarassingly parallel, as we will discuss below, and can be performed in modern datacenters in a reasonable time. The exact time depends on the clustering change, the overhead of communication between machines, the existing load of the datacenter, etc. For typical changes, for example, the sampling of items for exploring the impact can finish within 15 minutes, and the sampling of item pairs for quality metrics can finish within 45 minutes.

Suppose we want to sample elements from a population $P$, where each element $e$ has a weight $w_e$.
Conceptually, we can imagine elements being sampled as time progresses. The next draw of an element $e$ happens at a random point in time that is exponentially distributed with a rate parameter of $w_e$. Hence, we can associate each element $e$ with an initial draw time $dt_0(e) \sim \mathrm{Exp}(w_e)$. A sample \textit{without replacement} of $N$ elements is then simply the $N$ elements with the smallest initial draw times. To take a sample \textit{with replacement} of $N$ unique elements, we proceed as follows:
\begin{enumerate}
    \item Take the $N$ elements with the smallest initial draw times. Call this set $S$.
    \item Compute the maximum initial draw time $M$ among the elements in $S$. This is conceptually the point in time when the last draw happens.
    \item Associate each element $e$ in $S$ with a draw count $dc(e)$. To do that, we compute the time duration between its first draw and $M$ and denote it with $\mathit{td}(e) = M - dt_0(e)$. The element $e$ has already been drawn once, and in the remaining time duration $\mathit{td}(e)$, it is drawn a random number of times that is Poisson-distributed with parameter $w_e \cdot \mathit{td}(e)$. So the final draw count of element $e$ is given by $dc(e) \sim 1 + \mathrm{Pois}(w_e \cdot \mathit{td}(e))$. \label{laststepinsamplingwithreplacement}
\end{enumerate}

Alternatively, we can obtain a sample with replacement of $N^\prime$ unique elements, where $N^\prime \leq N$, by performing successive draws in step~\ref{laststepinsamplingwithreplacement}. That can be done as follows:
\begin{enumerate}[label=3.{\arabic*}]
    \item Associate each element in $S$ with its next draw time. Initially, the next draw time of element $e$ is simply $dt_0(e)$.
    \item Repeatedly draw the element $e$ with the smallest next draw time that is at most $M$. Upon drawing an element $e$ with the smallest next draw time $dt_i(e)$, compute a new next draw time for it, namely $dt_{i+1}(e) \sim dt_i(e) + \mathrm{Exp}(w_e)$.
\end{enumerate}

Creating a sample with replacement incrementally like that can be useful in ABCDE when there is a fixed budget for human judgements. We would like to fill the whole budget, yet many of the sampled item pairs won't count towards the budget. For example, the self-pairs $(i, i)$ don't require human judgement, and since the underlying equivalence relation is symmetric, we need only one human judgement when we sample both $(i, j)$ and $(j, i)$. Drawing item pairs one by one for the sample is helpful, because then we can continue until the budget is exactly filled. Another benefit of the approach is that we do not need to know the exact budget at the time when the population of item pairs is expanded. For example, if we know that the maximum budget is $B = 10000$ pairs, then we can obtain a set $S$ with $10 \cdot B = 100000$ pairs and persist it together with the pairs' weights and initial draw times. Obtaining a concrete sample is then fast when the actual budget becomes known later on.

\section{Estimating $\DeltaPrecision$ for slices of items}

The treatment in the main part of this paper focused on estimating the overall $\DeltaPrecision$ of the clustering change. As mentioned, humans can provide judgements for only a relatively small set of item pairs, and there are usually too few judgements to provide estimates about arbitrary slices of items with high confidence. 

In some cases it might be possible to obtain many judgements. For example, the evaluation might consult an ML model (such as a Large Language Model) instead of humans for judgements. The model might well be too expensive to produce judgements for hundreds of billions of pairs, but perhaps it can process several tens of thousands of pairs with a reasonable cost and latency.

If more judgements are available, then it might make sense to estimate the $\DeltaPrecision$ for slices of items. Two operations are of particular interest:
\begin{enumerate}
\item Estimate $\DeltaPrecision(I)$, the delta precision for a particular slice of items $I \subseteq T$. This is the expected value of $\DeltaPrecision(i)$ over $i \in I$.
\item Estimate $\DeltaPrecisionContribution(I)$, the contribution of a given slice $I \subseteq T$ to the overall $\DeltaPrecision(T)$. If several slices form a partitioning of $T$, then the sum of their contributions will be equal to the overall $\DeltaPrecision$.
\end{enumerate}

As will soon become clear, the two notions have a straightforward relationship: $\DeltaPrecisionContribution(I) = \frac{\weight(I)}{\weight(T)} \DeltaPrecision(I)$.

\subsection{Estimating $\DeltaPrecision(I)$}

Recall from before that
$$\DeltaPrecision(i) = \sum_{j \in \Base(i) \cup \Exp(i)} u_{j} l_{j} \indicate\left(i \match j\right)$$
For an arbitrary set of items $I \subseteq T$, the $\DeltaPrecision(I)$ is the weighted average:
\begin{align}
\DeltaPrecision(I) &= \frac{1}{\weight(I)} \sum_{i \in I} \weight(i) \cdot \DeltaPrecision(i) \\
&= \frac{1}{\weight(I)} \sum_{i \in I} \weight(i) \sum_{j \in \Base(i) \cup \Exp(i)} u_{j} l_{j} \indicate\left(i \match j\right) \\
&=  \sum_{i \in I} \sum_{j \in \Base(i) \cup \Exp(i)} \frac{\weight(i)}{\weight(I)}  u_{j} l_{j} \indicate\left(i \match j\right)
\end{align}
We can define
\begin{align}
v_{ij} = \frac{\weight(i)}{\weight(I)} u_{j} = \frac{\weight(i)}{\weight(I)} \left(u_{ij} \frac{\weight(T)}{\weight(i)}\right) = \frac{\weight(T)}{\weight(I)} u_{ij}
\end{align}
and since $l_j = l_{ij}$, we obtain
\begin{align}
\DeltaPrecision(I) = \sum_{i \in I} \sum_{j \in \Base(i) \cup \Exp(i)} v_{ij} l_{ij} \indicate\left(i \match j\right)
\end{align}

We can estimate this by sampling pairs $(i, j)$, where $i \in T$ and $j \in \Base(i) \cup \Exp(i)$, according to $v_{ij}$, computing the average of $l_{ij} \indicate\left(i \match j\right)$ for the sample, and then multiplying that by
\begin{align*}
&\ \ \ \ \sum_{i \in I} \sum_{j \in \Base(i) \cup \Exp(i)} v_{ij} \\
&= \frac{\weight(T)}{\weight(I)} \sum_{i \in I} \sum_{j \in \Base(i) \cup \Exp(i)} u_{ij} \\
&= \frac{\weight(T)}{\weight(I)} \pairweight(I)
\end{align*}
where the definition of $\pairweight(I)$ is obvious.

Note that sampling according to $v_{ij}$ is the same as sampling according to $u_{ij}$. So we can estimate $\DeltaPrecision(I)$ from the same sample that was used to estimate overall $\DeltaPrecision(T)$. We just need to track which item of the pair is the vantage point (e.g. by using labels or by always recording the vantage point $i$ as the first item of the pair). Then we can restrict the overall sample to the pairs whose vantage points are in $I$ and compute their average of $l_{ij} \indicate\left(i \match j\right)$; the multiplier must be computed separately, exactly or approximately, ahead of time or on the fly, but it does not involve human judgements.

Please keep in mind that some slices might not be well-represented in the overall sample, so dedicated samples might be needed if you really need reliable estimates for those.

\subsection{Estimating $\DeltaPrecisionContribution(I)$}

Recall from before that overall $\DeltaPrecision$ is equal to
$$\DeltaPrecision(T) = \sum_{i \in T} \sum_{j \in \Base(i) \cup \Exp(i)} u_{ij} l_{ij} \indicate(i \match j)$$
The contribution of an item slice $I$ to the overall $\DeltaPrecision$ is simply
\begin{align}
\DeltaPrecisionContribution(I) = \sum_{i \in I} \sum_{j \in \Base(i) \cup \Exp(i)} u_{ij} l_{ij} \indicate(i \match j)
\end{align}

If $T$ is partitioned into disjoint slices, say $T_1 \ldots T_n$, then the overall $\DeltaPrecision$ will be equal to the sum of the $\DeltaPrecisionContribution(T_s)$ for $s \in \{1, 2, \ldots, n\}$, so it is possible to understand/explain the overall $\DeltaPrecision$ in terms of the per-slice contributions.

We can estimate $\DeltaPrecisionContribution(I)$ directly: sample pairs $(i, j)$, where $i \in I$ and $j \in \Base(i) \cup \Exp(i)$, according to $u_{ij}$, compute the sample's average of $l_{ij}\indicate\left(i \match j\right)$
and multiply it by $\pairweight(I)$.

Alternatively, if we already/also have an estimate of $\DeltaPrecision(I)$, then we can just multiply it by $\frac{\weight(I)}{\weight(T)}$ because
\begin{align}
\DeltaPrecisionContribution(I) &= \sum_{i \in I} \sum_{j \in \Base(i) \cup \Exp(i)} u_{ij} l_{ij} \indicate(i \match j) \\
&= \sum_{i \in I} \sum_{j \in \Base(i) \cup \Exp(i)} \frac{\weight(I)}{\weight(T)} v_{ij} l_{ij} \indicate(i \match j) \\
&= \frac{\weight(I)}{\weight(T)} \sum_{i \in I} \sum_{j \in \Base(i) \cup \Exp(i)} v_{ij} l_{ij} \indicate(i \match j) \\
&=  \frac{\weight(I)}{\weight(T)} \DeltaPrecision(I)
\end{align}

\end{document}